\documentclass[fleqn,usenatbib]{mnras}

\usepackage{newtxtext,newtxmath}

\usepackage[T1]{fontenc}

\DeclareRobustCommand{\VAN}[3]{#2}
\let\VANthebibliography\thebibliography
\def\thebibliography{\DeclareRobustCommand{\VAN}[3]{##3}\VANthebibliography}

\usepackage{graphicx}	
\usepackage{amsmath}	

\usepackage{color}

\definecolor{auburn}{rgb}{0.43, 0.21, 0.1}

\title[Cores expected in dark matter numerical simulations]{Numerical simulations of dark matter haloes  produce polytropic central cores when reaching thermodynamic equilibrium}
\author[S\'anchez Almeida \& Trujillo]{
  Jorge S\'anchez Almeida$^{1,2}$\thanks{E-mail: jos@iac.es} 
and
Ignacio Trujillo$^{1,2}$
\\
$^1$Instituto de Astrof\'\i sica de Canarias, La Laguna, Tenerife, E-38200, Spain\\
$^2$Departamento de Astrof\'\i sica, Universidad de La Laguna, Spain
}

\date{Accepted XXX. Received YYY; in original form ZZZ}

\pubyear{2021}

\begin{document}
\label{firstpage}
\pagerange{\pageref{firstpage}--\pageref{lastpage}}
\maketitle

\begin{abstract}
Self-gravitating astronomical objects often show a central plateau in the density profile ({\em core}) whose physical origin is hotly debated.  Cores are theoretically expected in N-body systems of maximum entropy, however, they are not present in the canonical N-body numerical simulations of cold dark matter (CDM). Our work shows that despite this apparent contradiction between theory and numerical simulations, they are fully consistent. Simply put, cores are characteristic of systems in thermodynamic equilibrium, but thermalizing collisions are purposely suppressed in CDM simulations. When collisions are allowed, N-body numerical simulations develop cored density profiles, in perfect agreement with the theoretical expectation. We compare theory and two types of numerical simulations: 
(1) when DM particles are self-interacting (SIDM) with enough cross-section, then the effective two-body relaxation timescale becomes shorter than the Hubble time resulting in cored DM haloes. The haloes thus obtained,  with masses from dwarf galaxies to galaxy clusters, collapse to a single shape after normalization, and this shape agrees with the polytropic density profile theoretically expected.  
(2) The inner radii in canonical N-body numerical simulations are always discarded because the use of finite-mass DM particles artificially increases the two-body collision rate. We show that the discarded radii develop cores which are larger than the employed numerical softening and have polytropic shape independently of halo mass.
Our work suggests that the presence of cores in simulated (or observed) density profiles can used as evidence  for systems in thermodynamic equilibrium. 
\end{abstract}




\section{Introduction}\label{sect:intro}

The mass density profiles in the centers of dwarf galaxies are well reproduced by polytropes \citep{2020A&A...642L..14S}, which are theoretical profiles expected in N-body self-gravitating systems of maximum Tsallis entropy \citep{1993PhLA..174..384P,2005PhyA..350..303L}. The same solutions also explain the stellar surface density profiles observed in globular clusters  (Trujillo et al. 2021, in preparation). Polytropic DM haloes fit the velocity dispersion observed in many late-type galaxies \citep{2010MNRAS.405...77S}, reinforcing the practical interest of polytropes to describe real self-gravitating systems. Other astrophysical problems showing their validity are discussed in, e.g.,  \citet[][]{2013SSRv..175..183L} and \citet{2013ApJ...777...20S}.

Polytropes are characterized by a central {\em plateau} or {\em core}, i.e., 
\begin{equation}
  \frac{d\log\rho}{d\log r}\longrightarrow 0 {\rm ~~when~~} {r \longrightarrow 0},
  \label{eq:coredef}
\end{equation}
where $\rho(r)$ stands for the variation of the density with $r$, the distance to the center of the gravitational well.  This holds true independently of whether the classical Boltzmann-Gibbs entropy or the Tsallis entropy (for self gravitating systems)  is used to define the thermodynamical equilibrium (Section~\ref{sect:polytropes}). However, the N-body numerical simulations of CDM predict density distributions
\citep[NFW profiles or Einasto profiles; e.g.,][]{1997ApJ...490..493N,2004MNRAS.349.1039N,2019Galax...7...81Z}  with {\em cusps}, i.e., where
\begin{equation}
  \frac{d\log\rho}{d\log r}\longrightarrow -1\, (\not= 0) {\rm ~~when~~} {r \longrightarrow 0}.
  \label{eq:cusps}
\end{equation}
Thus, theoretical N-body systems of maximum entropy should have cores (Eq.~[\ref{eq:coredef}]) which, however, are not present in the N-body numerical simulations of CDM  (Eq.~[\ref{eq:cusps}]). Even if these two facts seem to be in contradiction, they are not. Our work is aimed at showing the consistency of the two results.

The paper is organized as follows: Section~\ref{sect:polytropes} explains why maximum entropy profiles are expected to have cores. Section~\ref{sect:SIDM} shows how DM-only numerical simulations produce haloes that have not reached thermodynamic equilibrium yet, even though they have been growing for as long as the age of the Universe.
Section  \ref{sect:SIDM} explains that the numerical simulations of {\em self-interacting} DM (SIDM) produce density profiles in thermodynamic equilibrium provided the SIDM cross section is large enough. A comparison between polytropes and density profiles produced in SIDM simulations is carried out in Sects.~\ref{sect:SIDMhalo1},  which allows us to conclude that polytropes reproduced SIDM profiles extremely well. Incidentally, we also show that the unresolved DM clump-clump collisions are insufficient to account for the SIDM cross-sections needed to reproduce the observed cores (App.~\ref{app:appb}). The artificial cores produced in CDM numerical simulation within the radius of convergence are also compliant with  polytropes and thermodynamical equilibrium   (Section~\ref{sect:wang}). Our conclusions are included in Section~\ref{sect:discussion} .

\section{Self gravitating systems of maximum entropy}\label{sect:polytropes}
Following the principles of statistical physics, any self gravitating system with many particles should tend to reach the most probable configuration and, thus, it should maximize the entropy \citep[e.g.,][]{2020arXiv201211709A}. Using the classical Boltzmann-Gibbs entropy to define such thermodynamical equilibrium leads to the so-called {isothermal sphere}, which is a density distribution with infinity mass and energy \citep[][see also below]{2008gady.book.....B,2008arXiv0812.2610P} thus disfavoring this maximum entropy solution. However, the long-range forces that characterize self-gravitating systems are neglected in the Boltzmann-Gibbs entropy. Mounting evidence suggests that systems with long-range interactions admit long-life meta-stable states described by a maximum $S_q$ Tsallis entropy  \cite[][and references therein]{1988JSP....52..479T,2009insm.book.....T}.   Observational evidence for the validity of the $S_q$ statistics has been found in connection with various astrophysical problems (Section~\ref{sect:intro}).

The maximization of the $S_q$ entropy of a Newtonian self-gravitating spherically-symmetric  N-body system, under the constraints imposed by the conservation of total mass and energy, leads to a polytropic distribution  \citep{1993PhLA..174..384P,2005PhyA..350..303L}.
A polytrope  of index $m$ is defined as the solution of the Lane-Emden equation for the (normalized) gravitational potential $\psi$  \citep{1967aits.book.....C,2008gady.book.....B},
\begin{equation}
  \frac{1}{s^2}\frac{d}{ds}\Big(s^2\frac{d\psi}{ds}\Big)=
  \begin{cases}
    -3\psi^m & \psi > 0,\\
    0 & \psi \le  0,
  \end{cases}
  \label{eq:lane_emden}
\end{equation}
with $\psi(0)=1$.
The symbol $s$ stands for the scaled distance in the 3D space and the mass volume density is recovered from $\psi$ as
 \begin{displaymath}
   \rho(r) = \rho(0)\,\psi(s)^m,
 \end{displaymath}
 \vskip -6mm
  \begin{equation}
      r = b s.
    \label{eq:densityle}
  \end{equation}
  The symbols $b$ and $\rho(0)$ are the two integration constants which, together with $m$, define the density profile.  
 In general, Eq.(\ref{eq:lane_emden}) has to be integrated numerically to get $\psi(s)$ and so $\rho(r)$.  
 However, one can show that when $r$ is small ($r\lesssim b$) the polytropes can be approximated as,
 \begin{equation}
   \rho(r)\simeq \rho(0)\,\Big[1-r^2/(2b^2)\Big]^m,
   \label{eq:central}
 \end{equation}
which automatically fulfills Eq.~(\ref{eq:coredef}) and so all polytropes have cores (even when $m\rightarrow\infty$). Three additional properties of the polytropes are important in the present context. First,  the range of physically sensible polytropic indexes is 
\begin{equation}
  3/2 \le m \le 5,
  \label{eq:nlimits}
 \end{equation}
 set because polytropes with $m\le 3/2$ are unstable or have infinite density  and those with $m > 5$ have infinite mass \citep{1993PhLA..174..384P,2008gady.book.....B}. Second, the limit $m\rightarrow\infty$ corresponds to the so-called {\em isothermal sphere}, which is the maximum entropy solution obtained when using the classical Boltzmann-Gibbs entropy. In this case, since $m>5$, the polytrope has infinity mass \citep[e.g.,][]{2008gady.book.....B}. Third, Eq.~(\ref{eq:central}) can be rewritten as
 \begin{equation}
   \frac{\rho(r)}{\rho_\alpha} \simeq
   \Big[1+\frac{\alpha}{2m}\Big(1-\frac{r^2}{r_\alpha^2}\Big)\Big]^m
   \simeq 1+\frac{\alpha}{2}\Big(1-\frac{r^2}{r_\alpha^2}\Big),
   \label{eq:central2}
 \end{equation}
 with $r_\alpha$ defined as
 \begin{equation}
   \frac{d\ln\rho}{d\ln r}(r_\alpha)=-\alpha,
   \label{eq:logder}
\end{equation}
so that $\rho_\alpha=\rho(r_\alpha)$. The second approximate identity in the right-hand side of Eq.~(\ref{eq:central2}), which holds for $\alpha \ll  2m$, indicates that after normalization  by $\rho_\alpha$ and $r_\alpha$, all polytropes collapse to a single shape independent of $m$. In other words, except for a trivial normalization, all the polytropes look the same in their cores, independently of the polytropic index $m$.

 \section{Collisions in self-gravitating systems}\label{sect:SIDM}

\begin{figure}
	\includegraphics[width=\columnwidth]{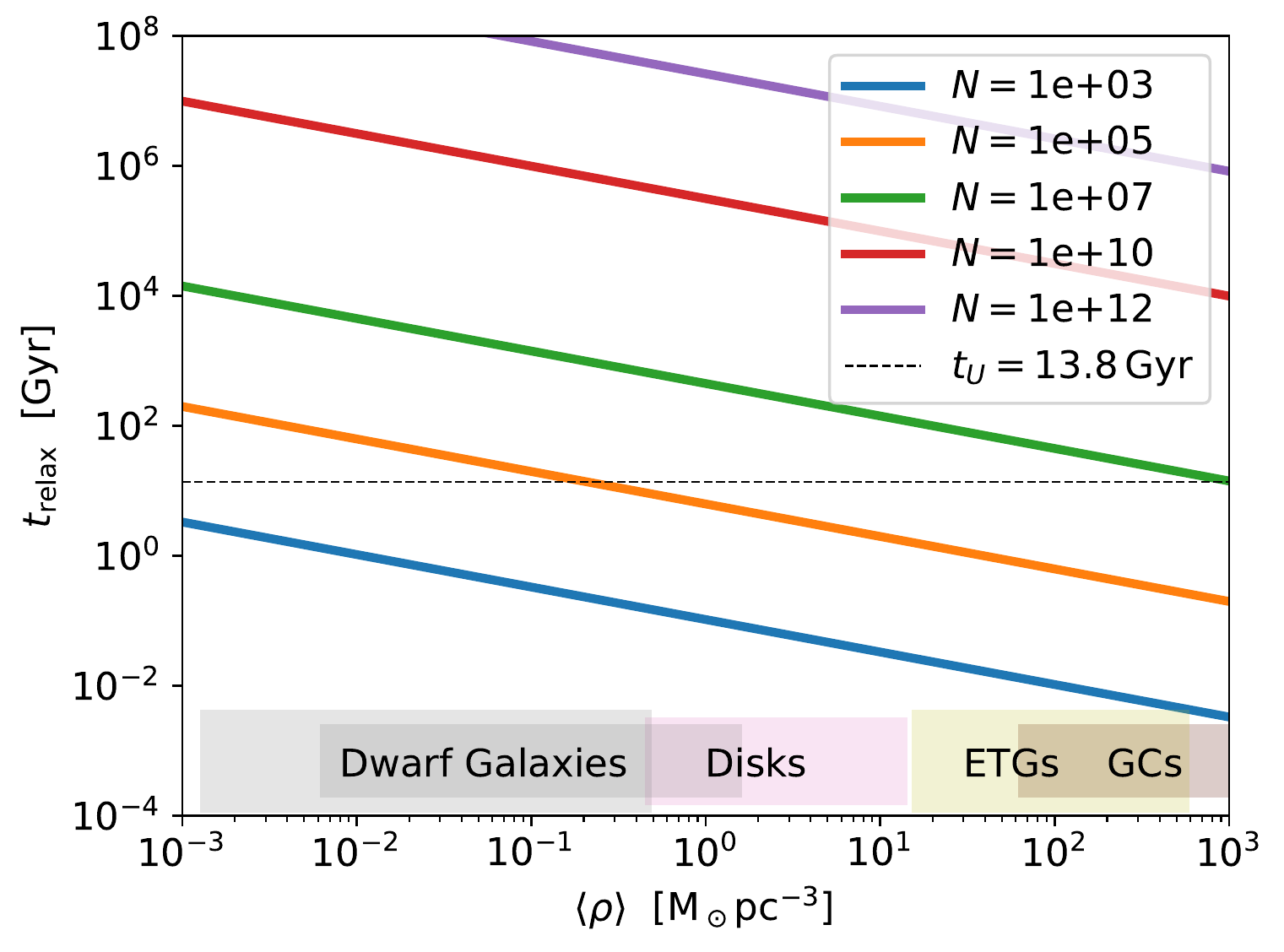}
        \caption{Two-body relaxation timescale versus mean density of the system. The dashed horizontal line shows the age of the Universe \citep[taken to be $\sim$13.8\,Gyr; ][]{2016A&A...594A..13P}, so that self-gravitating systems with parameters above this line have not reached thermodynamic equilibrium yet. The colored boxes at the bottom are shown for reference, and they indicate typical central densities in various astronomical self-gravitating systems: globular clusters \citep[GCs;][]{1996AJ....112.1487H}, early type massive galaxies \citep[ETGs; ][]{2014MNRAS.437.3670C}, galaxy disks \citep[Disks; ][]{2020MNRAS.493...87T}, and dwarf galaxies \citep[Dwarf Galaxies; ][]{2015AJ....149..180O,2020MNRAS.493...87T}.  GCs are stellar systems and so stellar densities are used. The central regions in Disks are dominated by baryons, therefore, we also show their stellar densities. ETGs have contribution from DM and stars, and total densities are shown. DM dominates the gravitational potential in Dwarf Galaxies, which appear in two boxes:  one including only stars (broader box) and another for stars and DM together (narrower box).
        }
    \label{fig:trelax}
\end{figure}

 In a system of $N$ particles moving under the influence of gravity, one can separate the force produced by the overall distribution of masses from the forces resulting from particle-particle encounters that perturb the motions created by the global gravitational potential. This second {\em two-body relaxation process} is very inefficient. The timescale for a particle to lose memory of its initial conditions, called {\em relaxation time}, $t_{\rm relax}$, is given by \citep[e.g.,][]{2008gady.book.....B},
 \begin{equation}
   t_{\rm relax}\simeq \frac{0.1\,N}{\ln N}\,t_{\rm cross},
   \label{eq:trelax}
   \end{equation}
   with $t_{\rm cross}$ the crossing time, i.e.,
   \begin{equation}
     t_{\rm cross}\simeq R/\upsilon\simeq \Big[\frac{4\pi}{3}G\,\langle\rho\rangle\Big]^{-1/2}, 
   \label{eq:tcross}
     \end{equation}
     set by the characteristic radius of the system, $R$, and the typical velocity of a particle in the self-gravitating system, $\upsilon$. The symbols $G$ and $\langle\rho\rangle$ stand for the gravitational constant and the mean density, respectively. Thus,
     \begin{equation}
       \langle\rho \rangle=  \frac{3N m_p}{4\pi R^{3}},
       \label{eq:rho1}
       \end{equation}
with $m_p$ the mass of the particle so that the total mass of the system is given by
       \begin{equation}
         M=N m_p.
       \label{eq:rho2}
         \end{equation}
         Figure~\ref{fig:trelax} shows $t_{\rm relax}$ for systems with densities in the observed range (between dwarf galaxies and globular clusters), and with a number of particles between $10^3$ and $10^{12}$. For $m_p = 1\,{\rm M}_\odot$, $M$ spans from $10^3$ and $10^{12}\,{\rm M}_\odot$. Note that for $N > 10^7$, even with the extreme densities existing at the centers of globular clusters, $t_{\rm relax}$ is longer than the age of the Universe ($t_U$), and the $N$-body system never loses memory of the initial conditions and thus cannot reach thermodynamic equilibrium. Only in GCs, where the mass particles are individual stars and $N \sim 10^5$, the two-body collisions operate in a timescale shorter than the age of the Universe (see Fig.~\ref{fig:trelax}).

The above conclusion is a well-stablished textbook result \citep[e.g.,][Section~1.2]{2008gady.book.....B} relevant in the present context. Many of the  CDM numerical simulations,  with an artificially small number of particles per DM halo, should be in thermodynamic equilibrium according to Fig.~\ref{fig:trelax}. However, the current numerical simulations try to capture the physics of {\em collision-less} systems. Thus they artificially avoid simulations where $t_{\rm relax} < t_U$. In other words, one of the criteria for the convergence of a simulation is that it must contain enough particles {\em so that the collisional relaxation time-scale is longer than the age of the Universe} \citep{2003MNRAS.338...14P}. The convergence radius of a CDM halo is also dictated by the same criteria: {\em convergence is obtained at radii that enclose a sufficient number of particles so that the local two-body relaxation time-scale is comparable to or longer than a Hubble time} \citep[][]{2019MNRAS.488.3663L}. In terms of Fig.~\ref{fig:trelax}, this implies that CDM numerical simulations with parameters below the horizontal dashed line are discarded. Thus, it is easy to understand why the profiles resulting from these numerical simulations seem to be set by the initial conditions during the Big Bang \citep{2014ApJ...790L..24C,2015ApJ...805L..16N,2020MNRAS.495.4994B}, rather than result from some kind of thermodynamic equilibrium. 
When the N-body simulations are allowed to evolve on timescales much longer than $t_{\rm relax}$, the resulting self-gravitating haloes develop central cores amenable to the polytropes introduced in Section~\ref{sect:polytropes} \citep[e.g.,][]{2003PhRvL..90r1101T}.

\begin{figure}
	\includegraphics[width=\columnwidth]{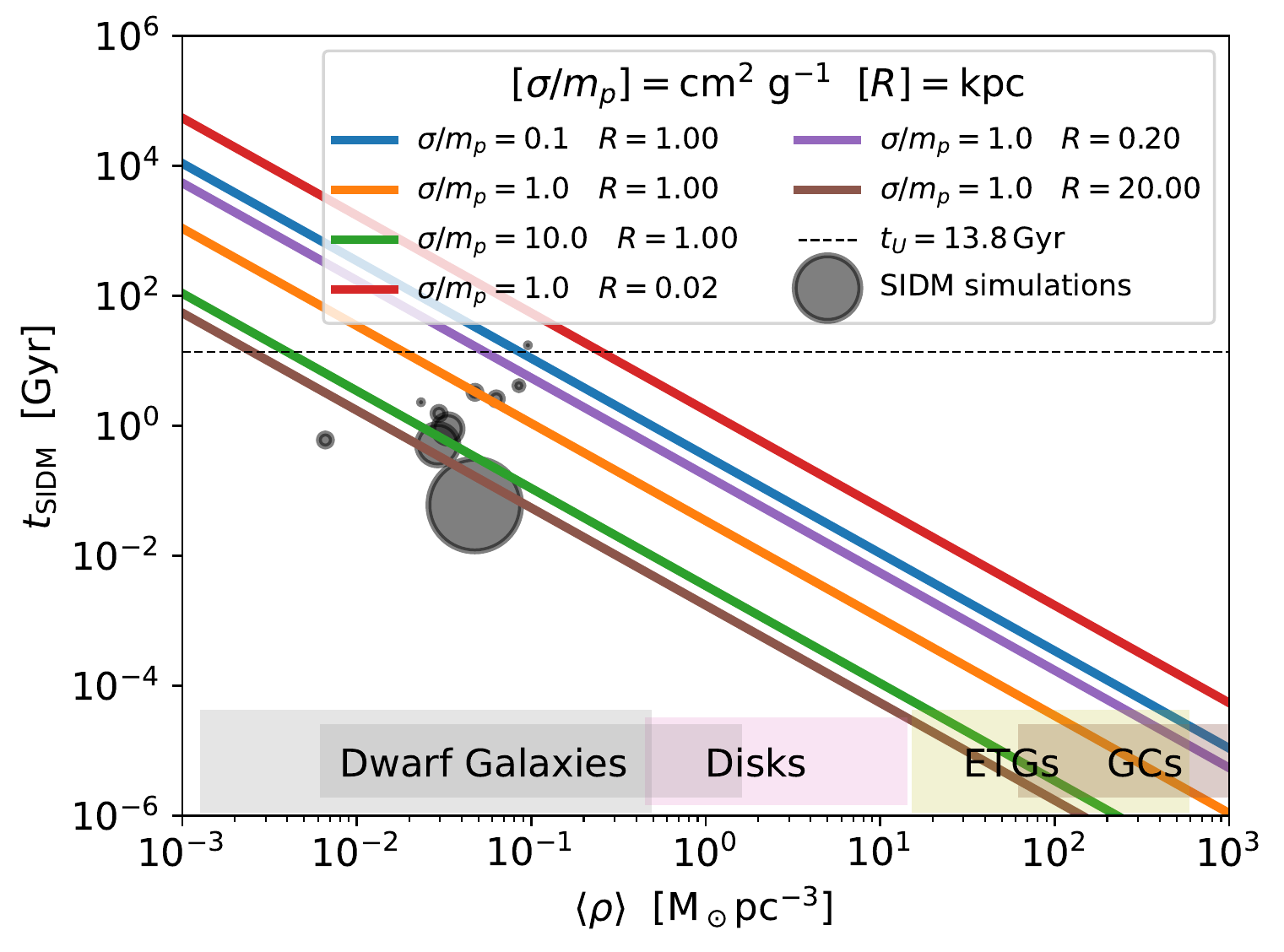}
        \caption{Timescale to collide assuming SIDM particles with cross-section per unit mass $\sigma/m_p$. Each line shows the variation with mean density $\langle\rho\rangle$ for systems with the same radius ($R$) and $\sigma/m_p$. In general, $t_{\rm SIDM}\ll t_U$ (the dotted line), and so we would expect the SIDM cosmological numerical simulations of galaxy formation to reach thermodynamic equilibrium. It has the same layout as Fig.~\ref{fig:trelax}. The boxes with typical central densities of various astronomical self-gravitating systems are the same as in Fig.~\ref{fig:trelax}, and we refer to the caption of this other figure for further details. The bullet symbols correspond to the central cores in the SIDM numerical simulations analyzed in Section~\ref{sect:SIDMhalo1}, with the symbol size scaling as $\sigma/m_p$ (50\,${\rm cm^2\,g^{-1}}$ largest  and  0.1\,${\rm cm^2\,g^{-1}}$ smallest). 
        }
    \label{fig:tSIDM}
\end{figure}

We will further elaborate on this in Section~\ref{sect:SIDMhalo1} using existing numerical simulations of SIDM, which can be employed in the present context with the arguments given below. 
%
%
The density profiles that arise from the customary CDM numerical simulations have cusps (Eq.~[\ref{eq:cusps}]), which represents a problem from the physical standpoint if CDM is to be responsible for the formation of galaxies since often galaxies show cores  (Eq.~[\ref{eq:coredef}]). One of the ways out of this so-called {\em core-cusp problem} \citep[e.g.,][]{2015PNAS..11212249W,2017Galax...5...17D} was assuming the DM particles to be self-interacting \citep[][]{2000PhRvL..84.3760S}. Hypothetical forces \citep[e.g.,][]{2018PhR...730....1T} add up to gravity allowing the DM particles to collide more often than the timescale given by Eq.~(\ref{eq:trelax}). In this case the relaxation timescale set by gravity has to be replaced with
\begin{equation}
  t_{\rm SIDM}\simeq \Big[\frac{\sigma}{m_p}\langle\rho\rangle\,\upsilon\Big]^{-1},
  \label{eq:sidm1}
\end{equation}
where $\sigma/m_p$ stands for the cross-section per unit mass of the self-interaction process \citep[e.g.,][]{2001ApJ...547..574D,2008ApJ...679.1173R,2013MNRAS.430...81R,2018JCAP...12..038S}. Equation~(\ref{eq:sidm1}) can be rewritten in a more intuitive way as
\begin{equation}
  t_{\rm SIDM}\simeq \Big[\frac{\sigma}{m_p}\langle\rho\rangle\,R\Big]^{-1}\,t_{\rm cross}\simeq \frac{\pi R^2}{N\sigma}\,t_{\rm cross},
  \label{eq:sidm2}
\end{equation}
showing $t_{\rm SIDM}$ to be formally identical to $t_{\rm relax}$ (Eq.~[\ref{eq:trelax}]) where the scaling $0.1 N/\ln N$ is replaced with the ratio between the geometrical area of the system ($\pi R^2$) and the total cross section ($N\sigma$). First, the new factor can be made very small provided $\sigma$ is large. Second, contrarily to Eq.~(\ref{eq:trelax}), the factor decreases with increasing $N$. Figure~\ref{fig:tSIDM} shows $t_{\rm SIDM}$ for standard values of $\sigma/m_p$ found in literature \citep[e.g.,][]{2015MNRAS.453...29E}. As soon as  $\sigma/m_p\gtrsim 1 {\rm cm}^2\,{\rm g}^{-1}$, it is easy to find systems for which $t_{\rm SIDM}\lesssim t_U$. Thus, SIDM numerical simulations with cross section around this value are expected to reach thermodynamic equilibrium and thus, according to the theory in Section~\ref{sect:polytropes}, to develop cores.  This is shown to be the case in Section~\ref{sect:SIDMhalo1}.

Equation (\ref{eq:sidm2}) gives  a relaxation timescale for the SIDM cross section. The exercise also works in the other direction, i.e., expressing the two-body relaxation timescale (Eq.~[\ref{eq:trelax}]) in terms of an effective cross section.  This exercise is particularly important in the context of CDM simulations since the CDM structures are hierarchical by nature so, for any mass resolution there is always a spectrum of unresolved DM clumps. Since clumps are far less numerous than DM particles, the relaxation timescale associated with clump-clump collisions could be short.  The question arises as to whether two-body collisions between unresolved clumps could thermalize the self-gravitating system. Assuming the clumps to be point masses, we work out in Appendix~\ref{app:appb} the effective cross section resulting from unresolved clump-clump collisions. The model depends on five free parameters, namely, the total mass and size of the system, the mass resolution of the numerical simulation ($m_{\rm resol}$), the mass of the DM particle, and the exponent of the power law characterizing the mass function of the unresolved clumps ($-\beta$).  We explore a reasonable range of parameters in Appendix~\ref{app:appb}, namely, $6\le \log(M/{\rm M}_\odot)\le 12$, 10\,pc\,$\le R \le$\,10\,kpc,  $1\le \log(m_{\rm resol}/{\rm M}_\odot)\le 5$, and $0.5\le \beta \le 2.5$. Within this range of $\beta$, the true DM particle mass $m_p$ plays no significant role and was set to $10^{-10}\, {\rm M}_\odot$. The calculation yields an effective cross section  $< 1\, {\rm cm}^2\,{\rm g}^{-1}$, which is insufficient to create cores in $t_U$ as discussed above and  in Section~\ref{sect:SIDMhalo1}.


\section{Comparison between polytropes and simulated SIDM haloes}\label{sect:SIDMhalo1}

\citet{2015MNRAS.453...29E} carry out a numerical simulation of SIDM haloes corresponding to dwarf galaxies ($5 < \log [M_\star/{\rm M}_\odot]  < 7$), and one of their key conclusions was that cores appear without fine-tuning the cross section. This conclusion fits well with the argument given in Section~\ref{sect:SIDM}. The cross section  has to be just large enough for the N-body system to settle into thermodynamic equilibrium during the time-span of the simulation. Provided the condition is met, the resulting profile is insensitive to the actual value of the cross section.  \citet{2015MNRAS.453...29E} explain how the scattering cross section produces haloes if it is large enough to yield a scattering per particle during the age of the system, very much in line with the arguments defended in Section~\ref{sect:SIDM}. Other authors also provide similar explanations for the emergence of cores in SIDM simulations \citep[e.g.,][]{2015MNRAS.452.1468F,2018JCAP...12..038S}.

\begin{figure*}
  \includegraphics[width=0.89\columnwidth]{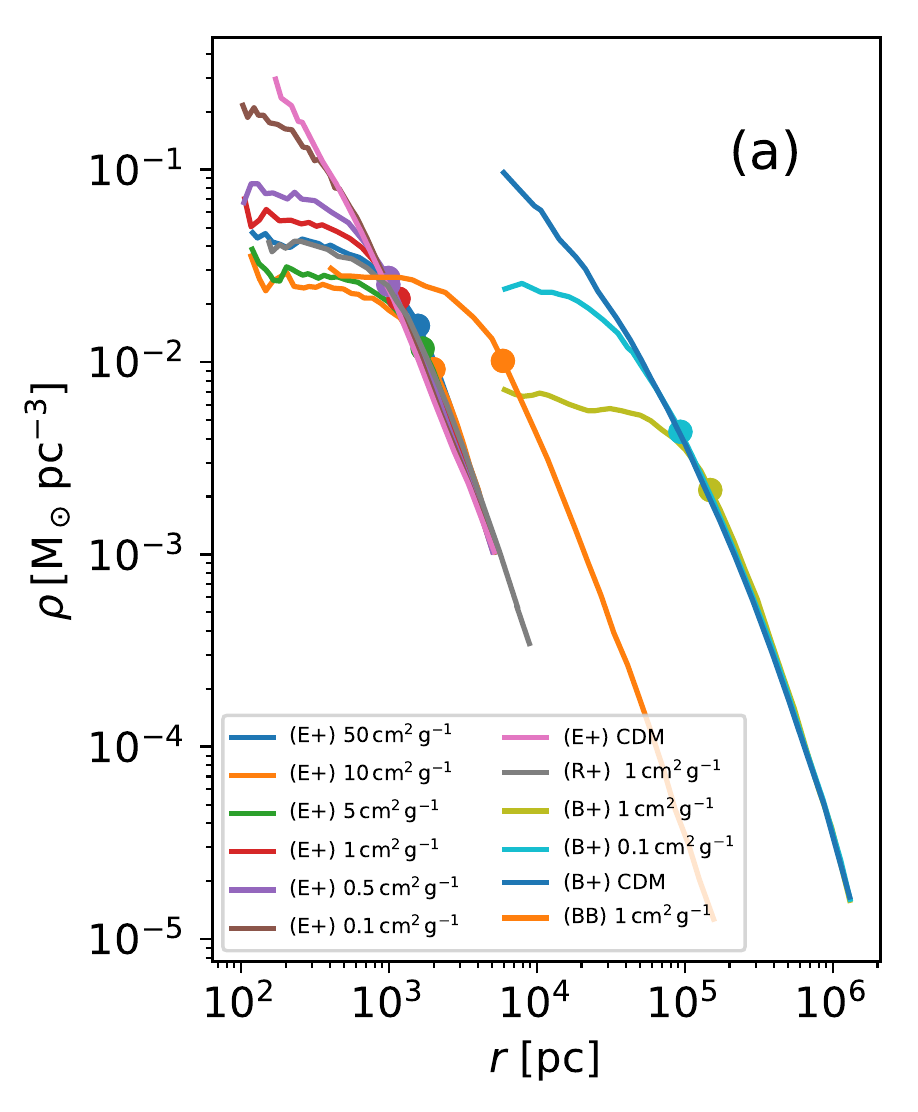}
  \includegraphics[width=0.9\columnwidth]{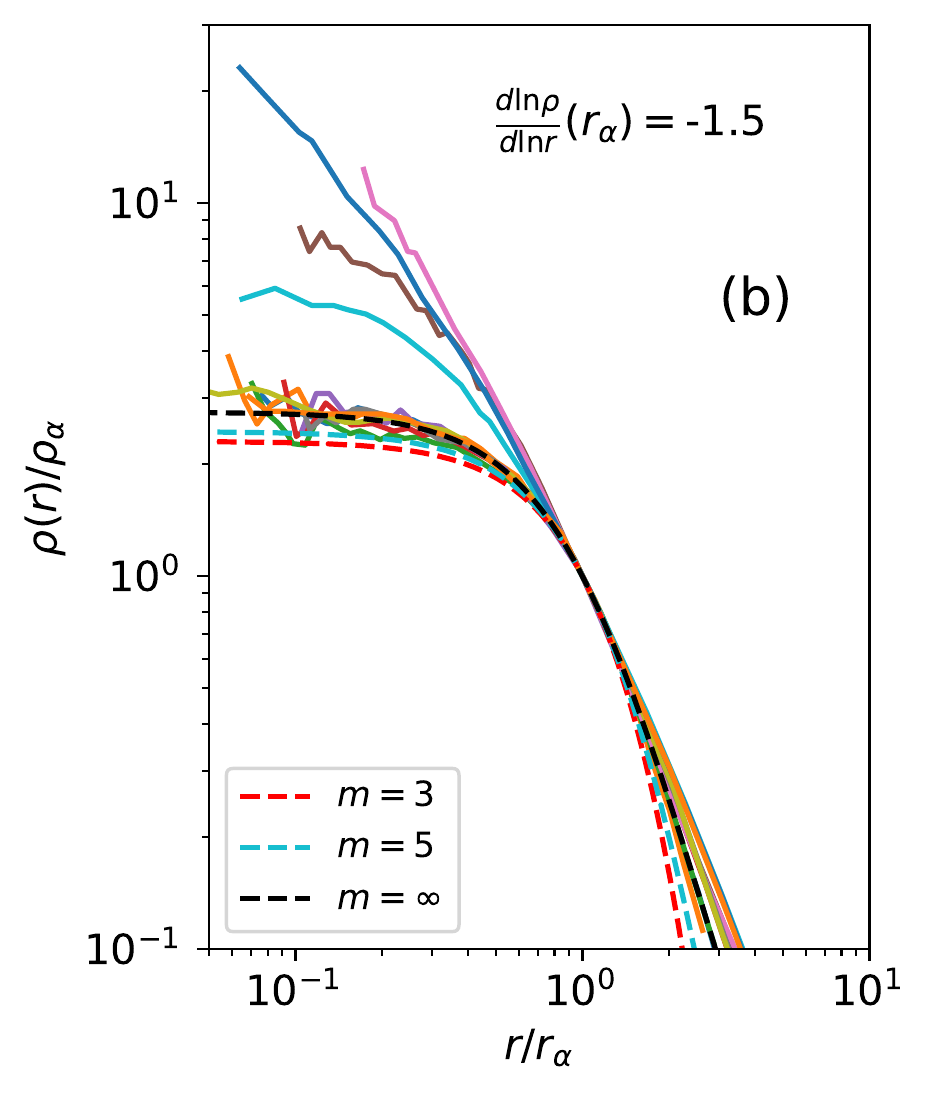}
  \includegraphics[width=0.73\columnwidth]{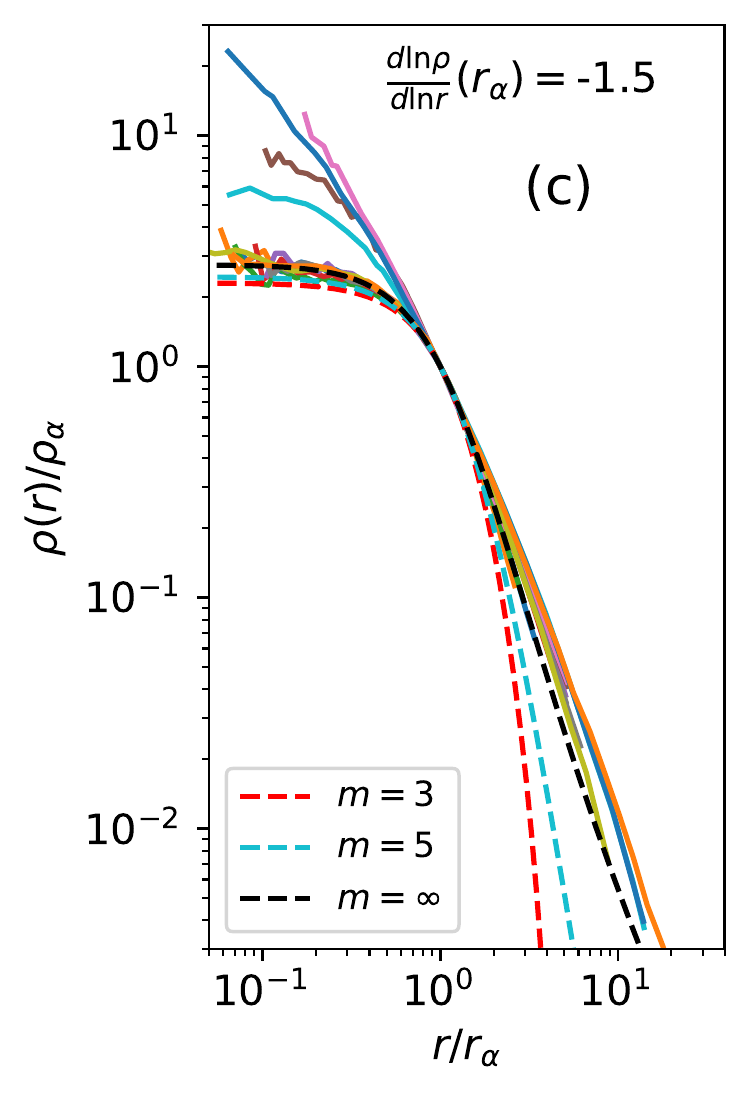}
  \includegraphics[width=0.9\columnwidth]{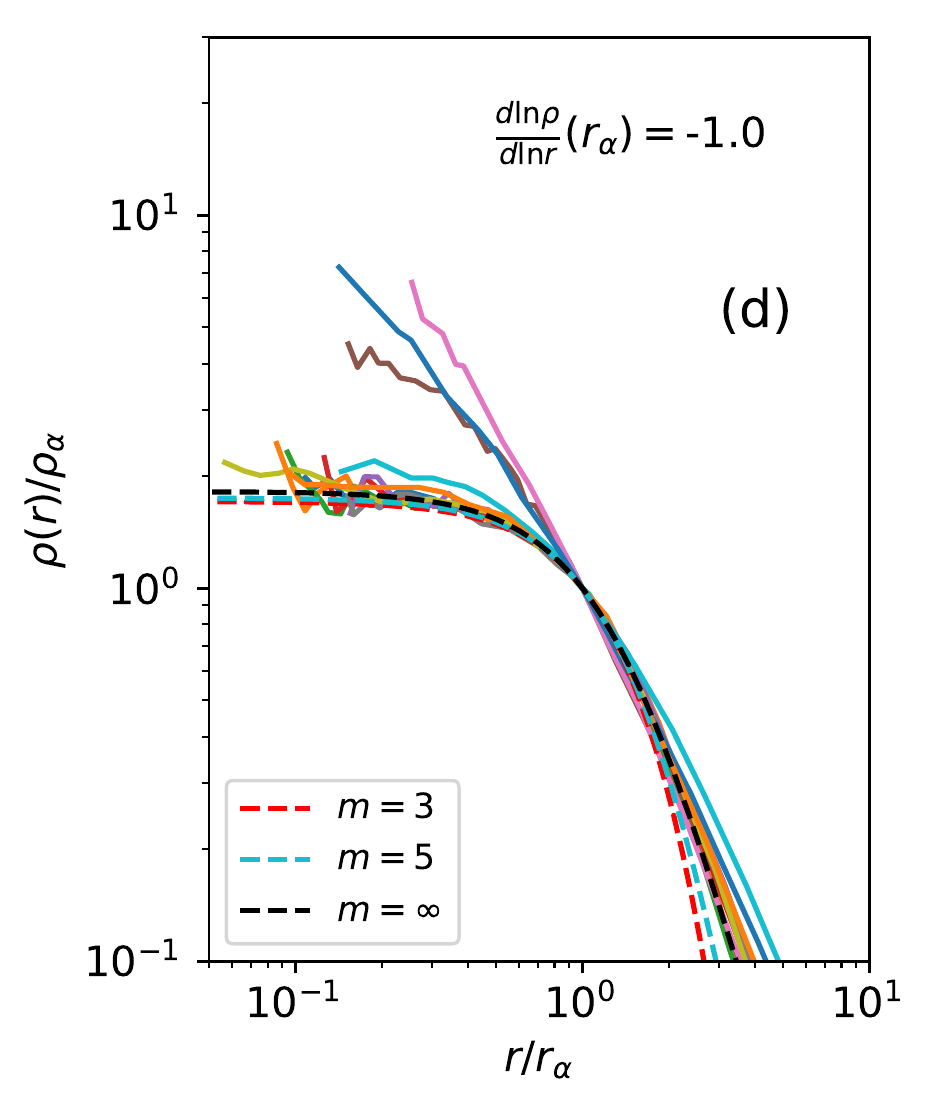}
  \caption{(a) Mass density profiles produced by a number of SIDM numerical simulations: \citet{2015MNRAS.453...29E} (E+), \citet{2017ARA&A..55..343B} (BB), \citet{2017MNRAS.472.2945R} (R+), and \citet{2018MNRAS.474..746B} (B+). They encompass halo masses going from dwarf galaxies (E+, R+) to clusters of galaxies(B+), including a MW halo (BB).  The $\sigma/m_p$ used in the simulation appears in the inset, with CDM standing for $\sigma/m_p=0$.
    (b) Profiles shown in (a) normalized to the radius and density of the point where the logarithmic derivative ($-\alpha$)  reaches a particular value ($-1.5$ in this case). The pairs $(\rho_\alpha,r_\alpha)$ are shown as bullet symbols in (a). As soon as $\sigma/m_p\gtrsim 1\,{\rm cm^2}\,{\rm g}^{-1}$, the numerical simulations develop a core, which is very similar to the core predicted by the polytropes. The shape of this core is very insensitive to the polytropic index $m$ ($3$, $5$, and $\infty$ are included in the plot).
    (c) Same as (b) with the axes expanded so as to see the tails of the density distribution. Polytropes do not fit the outskirts well except, perhaps, $m\longrightarrow\infty$. 
(d) Same as (b) with $\alpha=1$.
  }
    \label{fig:sidmhaloes}
\end{figure*}

Thus, SIDM N-body numerical simulations reduce the relaxation timescale allowing the systems to reach thermodynamical equilibrium within the time-span of the simulation.  Here we provide quantitative arguments supporting this view. Figure~\ref{fig:sidmhaloes} shows density profiles of four different SIDM numerical simulations. They go from galaxy cluster mass haloes \citep{2018MNRAS.474..746B},  to dwarf mass haloes \citep{2015MNRAS.453...29E,2017MNRAS.472.2945R}, including a MW mass halo \citep{2017ARA&A..55..343B}. The diversity of masses, sizes and central densities are shown in Fig.~\ref{fig:sidmhaloes}a. The range of $\sigma/m_p$ spans from zero (truly CDM) to $50\,{\rm cm^2}\,{\rm g}^{-1}$. These very different profiles collapse to a narrow range of shapes if they are normalized to a  radius ($r_\alpha$) and a density ($\rho_\alpha$) and at which the logarithmic slope of the density has a fixed value ($-\alpha$; defined in Eq.~[\ref{eq:logder}]).
Figure~\ref{fig:sidmhaloes}b shows the profiles in Fig.~\ref{fig:sidmhaloes}a normalized for $\alpha=1.5$. As soon as $\sigma/m_p\gtrsim 1\,{\rm g}\,{\rm cm}^{-2}$, all the profiles show a very similar core, which also agrees with the cores predicted by the polytropes. Figure~\ref{fig:sidmhaloes}b includes three polytropes; $m=3$, $5$, and $\infty$. The agreement between SIDM numerical simulations and polytropes is excellent, which is particularly revealing since no fitting is involved in the comparison between them. Note that the cores corresponding to very different polytropic indexes are hard to distinguish, as expected (Eq.~[\ref{eq:central2}]). 
Figure \ref{fig:sidmhaloes}c is identical to \ref{fig:sidmhaloes}b except that the axes have been expanded to see the tail of the density profiles. It is clear that the simulations and the polytropes separate when the radius increases  (except for $m\rightarrow\infty$) which can be understood in terms of the increase in the relaxation time due to the drop in density (see the discussion below and Section~\ref{sect:discussion}).    
Figure~\ref{fig:sidmhaloes}d is similar to \ref{fig:sidmhaloes}b except that $\alpha = 1$, and it has been included to show how that the qualitative behavior is insensitive to the actual $\alpha$.

The relaxation timescale corresponding to the central cores of the above simulations are included in Fig.~\ref{fig:tSIDM} (the bullet symbols). We have used Eq.~(\ref{eq:sidm2}) with the central densities and the core radii shown in Fig.~\ref{fig:sidmhaloes}a ($\sim2\,\rho_\alpha$ and $r_\alpha$, respectively). Obviously, CDM simulations have $t_{\rm SIDM}=\infty$ , and do not appear in the plot. All simulations (but one) present timescales shorter than the age of the universe, and so the central parts are expected to be thermalized (as suggested in the references given above).  The only simulation with $t_{\rm SIDM} >   t_U$ is the one with $\sigma/m_p=0.1\,{\rm cm}^2\,{\rm g}^{-1}$ in dwarf galaxies \citep{2015MNRAS.453...29E}, which do not present an obvious core (Figs.~\ref{fig:sidmhaloes}b and \ref{fig:sidmhaloes}d). Keeping everything else the same, a decrease in density produces an increase in  $t_{\rm SIDM}$ following the straight lines in Fig.~\ref{fig:tSIDM}. Therefore, if the density drops by one order of magnitude with respect to the central densities, then most symbols will move above $t_U$ in Fig.~\ref{fig:tSIDM}. Thus, given the densities in the outskirts of the simulated haloes, we do not expect them to have reached thermodynamical equilibrium yet. This may explain deviations from polytropes in the outskirts together with the  (heuristically found)  piecewise profiles discussed in Section~\ref{sect:SIDMhalo2}.

Several conclusions can be drawn from Fig.~\ref{fig:sidmhaloes}: (1) different SIDM numerical simulations produce density profiles that are very similar to one another once they are normalized. This is true even when comparing dwarf galaxy haloes with galaxy cluster haloes. (2) SIDM numerical simulations produce central cores very similar to the cores predicted by the thermodynamic equilibrium, i.e., polytropes. (3) Cores appear as soon as $\sigma/m_p\gtrsim 1$\,cm$^{2}$\,g$^{-1}$ \citep[][]{2015MNRAS.453...29E}. (4) The low densities in the outskirts of the simulations prevent them from being in thermodynamic equilibrium.


\subsection{Comparison with isothermal Jeans modeling profiles}\label{sect:SIDMhalo2}

\citet{2020arXiv200907844R} convincingly show how the radial average of the density profiles arising from SIDM numerical simulations \citep[from][]{2019MNRAS.488.3646R} are accurately reproduced by profiles from {\em isothermal Jeans modeling}. These profiles are the combination of an isothermal sphere when $r < r_1$ and a NFW profile (Navarro, Frenk, and White \citeyear{1997ApJ...490..493N}) when $r > r_1$. The parameter $r_1$ is chosen as the radius where $t_{\rm SIDM}$ (in Eq.~[\ref{eq:sidm1}]) equals the age of the halo, taken to have an average value of  7.5\,Gyr. The underlying idea is that for $r < r_1$ the halo is in thermal equilibrium whereas it is not for  $r > r_1$. The mach between the two profiles is forced to be smooth requesting that the inner mass of the two profiles is the same,
\begin{equation}
  \int_0^{r_1}\,r^2\rho_{\rm NFW}(r)\,dr = \int_0^{r_1}\,r^2\rho_{\rm iso}(r)\,dr, 
\end{equation}
and the matched density profile is continuous at $r=r_1$,
\begin{equation}
  \rho_{\rm NFW}(r_1) = \rho_{\rm iso}(r_1).
\label{eq:condition1}  
  \end{equation}
  The symbols $\rho_{\rm NFW}$ and $\rho_{\rm iso}$ stand for the density of the NFW and the isothermal profile, respectively, explicitly,
  \begin{equation}
    \rho_{\rm NFW}(r)=\frac{4\,\rho(r_s)}{(r/r_s)\,(1+r/r_s)^2},
    \label{eq:condition2}
  \end{equation}
  and $\rho_{\rm iso}$ is the limit polytrope when $m\longrightarrow \infty$. The symbols $r_s$ is the so-called scale radius and $\rho(r_s)$ is a multiplicative factor.

\begin{figure}
  \centering
    \includegraphics[width=0.9\columnwidth]{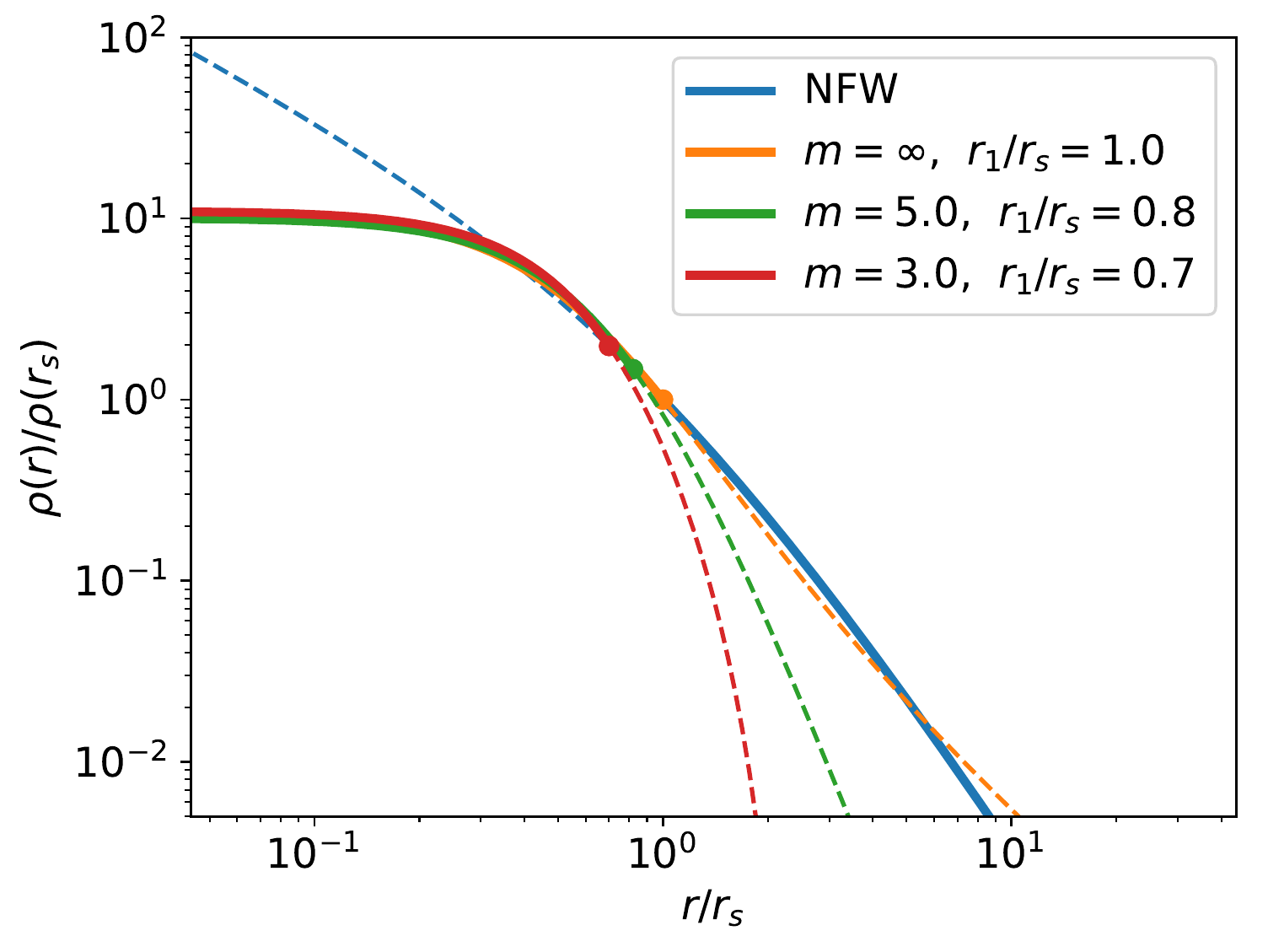}
  \caption{Isothermal Jeans model used by \citet{2020arXiv200907844R} -- an isothermal sphere for $r < r_1$ (the orange solid line; $m=\infty$) and a NFW profile for $r \geq r_1$ (the blue solid line). Dashed lines are used to show the profiles outside the range employed in the isothermal Jeans model. Virtually the same profiles are obtained when the isothermal sphere is replaced with another polytrope of any index. Examples corresponding $m=5$ and $m=3$ are included. The bullet symbols represent $r_1/r_s$, i.e.,  the point where the inner polytrope is forced to match the outer NFW profiles by fulfilling Eqs.(\ref{eq:condition1}) and (\ref{eq:condition2}). We employ $r_1/r_s\sim 1$, and the range of ordinates and abscissae in this figure mimics those used by \citet{2020arXiv200907844R}. Plots (not shown) for $r_1/r_s$ going from 0.5 to 2 are very similar.}
  \label{fig:robertson1}
  \end{figure}

The fact that the simulated SIDM profiles have this central thermal equilibrium core and an out-of equilibrium tail is very much in line with the arguments given in Section~\ref{sect:SIDMhalo1}, which reinforces them. Moreover, the selection of the isothermal equilibrium to represent the central core is arbitrary. Any other polytropes would do the work as well. Figure~\ref{fig:robertson1} shows examples of {\em isothermal Jeans models} similar to those used by \citet{2020arXiv200907844R}. The only free parameters is $r_1$. This parameter is not far from $r_s$ in the profiles shown by \citet{2020arXiv200907844R}. It is clear from Fig.~\ref{fig:robertson1} that profiles very similar to the isothermal Jeans models are obtained using virtually any value of $m$. As we prove in Section~\ref{sect:polytropes}, this is to be expected since the cores of all polytropes are alike.
  


  \section{Comparison between polytropes and CDM profiles within the convergence radius}\label{sect:wang}

\begin{figure}
  \includegraphics[width=0.9\columnwidth]{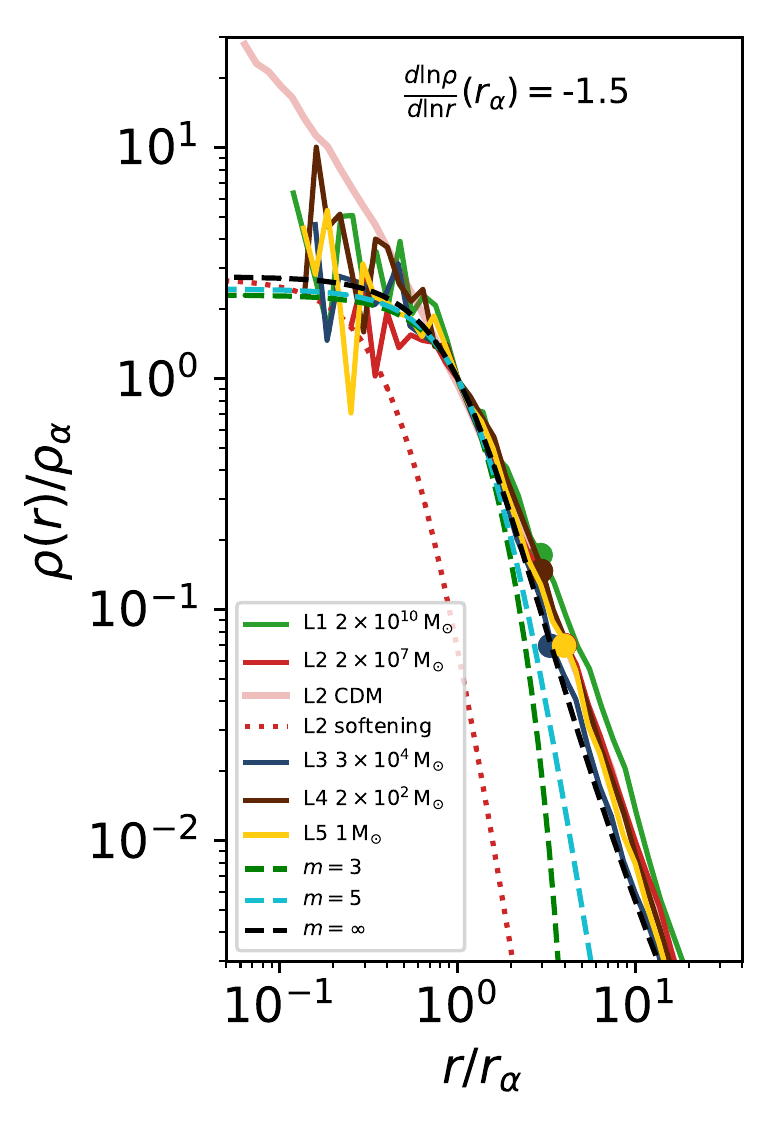}
  \caption{Density profiles for CDM numerical simulations showing the {\em artificial} cores appearing within the convergence radius (marked by color symbols), where the timescale for two-body collisions between the (artificially massive) DM particles used in the simulation is shorter than the age of the Universe. The profiles have been taken from \citet{2020Natur.585...39W}, and they represent DM haloes with masses differing by as much as 10 orders of magnitude (see the inset). The figure has the same layout as Fig.~\ref{fig:sidmhaloes}c, including the normalization of the haloes. The dashed lines represents polytropes, with the index included in the inset. The dotted line shows to core expected from the use of a finite softening length in the numerical scheme of the halo named L2. (Those for the others have not been included to avoid cluttering, but they are similar.) The pink solid line represents a properly resolved CDM halo corresponding to L2, and it closely follows a NFW profile.}
  \label{fig:wang20_dmhalosa}
\end{figure}

  The density profiles in CDM numerical simulations also show  cores within the convergence radius,  where the timescale for two-body collisions between the (artificially massive) DM particles used in the simulation is shorter than the age of the Universe (see Section~\ref{sect:SIDM}).
Within the convergence radius, the shape of the simulated DM haloes is an artifact created by the finite number of particles in the simulation \citep[e.g.,][]{2003MNRAS.338...14P,2019MNRAS.488.3663L} and, consequently, cores are disregarded for the usual analysis. However, since their shape is set by collisions, they can be used to test whether simulations reaching thermodynamic equilibrium have polytropic cores.
Figure~\ref{fig:wang20_dmhalosa} is a recent example taken from \citet{2020Natur.585...39W}.  They represent DM haloes with masses differing by as much as 10 orders of magnitude. The figure has the same layout as Fig.~\ref{fig:sidmhaloes}c, including the normalization of the haloes. The dashed lines show three representative polytropes, with the index labelled in the inset. The radius of convergence of the numerical simulations is indicated by color bullet symbols. Within the convergence radius, the simulated profiles are well reproduced by polytropes. These cores are very similar to the cores obtained from SIDM numerical simulations analyzed in Section~\ref{sect:SIDMhalo1} (the polytropes included in the figure are identical to those in  Fig.~\ref{fig:sidmhaloes}c and can be used as a reference to compare Figs.~\ref{fig:sidmhaloes}c and  \ref{fig:wang20_dmhalosa}).

The question arises as to whether these cores are artifacts produced by the use of a softening length in the numerical scheme \citep[e.g.,][]{2003MNRAS.338...14P}.  We think that this explanation can be ruled out for a number of reasons. First, 
the softening employed in the simulations is significantly smaller than the size of the resulting cores. This can be shown as follows: the softening is introduced to avoid artificially large two-body collisions. In practice, each particle is no longer a point mass but assumed to produce a force created by a Plummer potential \citep[e.g.,][]{1993ApJ...409...60D},   
\begin{equation}
    \psi(s)=\frac{1}{\sqrt{1+s^2}},
  \end{equation}
  which is just the polytrope of index $m=5$. Therefore, the density of the particles resulting from this potential is given Eq.~(\ref{eq:densityle}) with $b$ the softening length. Figure~\ref{fig:wang20_dmhalosa}, the red dotted line, shows the density corresponding to the softening length used in the halo named L2 (the red solid line).  The true core is around three times the core associated with gravitational softening kernel. Thus, this large difference suggests that the use of a integration scheme with finite softening length is not causing the cores.

Second, the fact that the softening length is not determining the size of the cores is implicit in the study carried out by  \citet[][]{2003MNRAS.338...14P}, and summarized in their Fig.~9. When the softening length tends to zero the profiles are independent of the softening, and differ from a pure NFW profile.
Finally, \citet{2019MNRAS.488.3663L} show how the convergence radius is independent of the softening when the softening length is significantly smaller than the convergence radius, which is case of the  simulations by \citet{2020Natur.585...39W} (the employed ratio is  $\sim0.15$). Thus, the softening could have been made arbitrarily small without changing the convergence radius, that is to say, without changing the core since these two length scales are related almost one-to-one. The ratio between size of the core and convergence radius is almost the same in all five profiles represented in Fig.~\ref{fig:wang20_dmhalosa}, despite the physical size of their cores differs by three orders of magnitude.

  \section{Discussion and conclusions}\label{sect:discussion}

  Many self-gravitating astronomical objects show cores whose origin is still debated in the literature (Section~\ref{sect:intro}, and references therein). The present work should be framed in this context, namely, as an exercise to clarify why cores are theoretically predicted in thermodynamic equilibrium but difficult to disclose in numerical simulations.  Specifically, theoretical N-body systems of maximum entropy should have cores (Eq.~[\ref{eq:coredef}]) which, however, are not present in the canonical N-body numerical simulations of CDM  (Eq.~[\ref{eq:cusps}]). These two facts are not in contradiction, though. Simply put, cores are characteristic of density profiles in thermodynamic equilibrium, but thermalizing collisions are purposely suppressed in CDM simulations. When enough collisions are allowed in CDM simulations, they develop density profiles with cores very much in agreement with the theoretical expectations.  

  The thermodynamic equilibrium of self-gravitating systems produce polytropes (see Section~\ref{sect:polytropes} for details and references). Polytropes depend on three parameters, namely, a two global scaling factors (in abscissae and ordinates) plus the index $m$,  which accounts for the overall shape. As we show in Section~\ref{sect:polytropes}, except for a trivial normalization, all  polytropes look the same in their cores, independently of $m$. This property is commonly observed in astronomical objects \citep[e.g., in dwarf galaxies;][]{2015AJ....149..180O}. In {\em collision-less} self gravitating systems, the only way to thermalization is via two-body collisions, which are extremely inefficient if the number of particles in the system is large (Section~\ref{sect:SIDM}), as expected in astronomical objects formed by myriads of DM particles. This is the reason why two convergence criteria are applied to the CDM numerical simulations to declare their results reliable \citep[e.g.,][]{2003MNRAS.338...14P}. First, the finite mass of the DM particles used in the simulation artificially shortens the two-body relaxation timescale, and this artificially-set timescale is asked to be longer than the age of the Universe, $t_U$. This requirement prevents the thermalization of the resulting DM haloes. Second, the artificial thermalization starts in the center of the DM haloes so the simulated centers are discarded within the so-called convergence radius. However, when these two restrictions are relaxed then cores with polytropic shape appear in the DM haloes produced in the numerical simulation. In the paper, we analyze two types of numerical simulations showing such cores.

  If the DM is assumed to be self-interacting (SIDM) then the effective relaxation timescale $t_{\rm SIDM}$ is shortened by an amount that depends on the SIDM cross section. As soon as  $\sigma/m_p\gtrsim 1 {\rm cm}^2\,{\rm g}^{-1}$, $t_{\rm SIDM}\lesssim t_U$ and the simulations produce cores. We show in   Section~\ref{sect:SIDMhalo1} how different SIDM from several authors produce cores which after normalization are very similar and equal to the core of a polytrope within error bars. The simulations comprise haloes that go all the way from dwarf galaxies to galaxy clusters. This agreement between simulated  haloes and polytropes holds even when baryons are included in the simulation (Section~\ref{sect:SIDMhalo2}).  

  If standard  CDM numerical simulations are analyzed within the convergence radius, they also show cores \citep[e.g.,][]{2020Natur.585...39W}. These cores are also indistinguishable from polytropes, a result which holds even with halo masses differing by as much as 10 orders of magnitude (Section~\ref{sect:wang}). Outside the convergence radius the profiles follow the canonical NFW or Einasto shape. Thus, because the profiles within the convergence radii are excluded from the standard analysis,  it is easy to understand why the profiles resulting from these numerical simulations seem to be set by the initial conditions during the Big Bang  \citep{2014ApJ...790L..24C,2015ApJ...805L..16N,2020MNRAS.495.4994B}.

  As a by-product of our work, we have shown how unresolved DM clump-clump collisions cannot thermalize the DM haloes within a timescale comparable to the age of the Universe (App.~\ref{app:appb}).

We conclude that the presence of cores in simulated or observed density profiles can used as evidence  for systems in thermodynamic equilibrium. Cores are theoretically predicted and also generated in self-gravitating system known to be in  thermodynamic equilibrium.
  



%
\section*{Acknowledgements}

Thanks are due to Angel R. Plastino for advice and support during the early development of this work, and for a careful reading of the original manuscript. Thanks area also due to Claudio Dalla Vecchia for insightful discussions and references, and to
Jie Wang for providing the data used in Fig.~\ref{fig:wang20_dmhalosa}.
  Our work makes use of various {\sc python} public packages including {\sc numpy} \citep{2011CSE....13b..22V}, {\sc matplotlib} \citep{2007CSE.....9...90H} , and  {\sc scipy} \citep{Jones01scipy}.
JSA acknowledges support  from the Spanish Ministry of  Science and Innovation, project PID2019-107408GB-C43 (ESTALLIDOS7),
and from the Gobierno de Canarias through EU FEDER funding,  I+D project PID2020010050.
IT acknowledges support from grant PID2019-107427GB-C32 from The Spanish Ministry of Science and Innovation, and also from the European Union's Horizon 2020 research and innovation programme under Marie Sk\l odowska-Curie grant agreement No 721463 to the SUNDIAL ITN network, and the European Regional Development Fund (FEDER), from IAC project P/300624, financed by the Ministry of Science, Innovation and Universities, through the State Budget and by the Canary Islands Department of Economy, Knowledge and Employment, through the Regional Budget of the Autonomous Community.

\section*{Data Availability}


  All the numerical simulations used in the work have been alredy published by other authors and so they are already publicly available. The Phyton routines using for plotting are also available upon request.







\appendix


\section{Two-body collision cross-section for a continuous distribution of unresolved DM clumps}\label{app:appb}

Starting from the primordial density fluctuations, the gravitational collapse produces a full distribution of DM haloes with masses going from galaxy clusters to Earth-mass clumps and even smaller \citep[e.g.,][]{2005Natur.433..389D,2020Natur.585...39W}. However, numerical simulations cannot treat all these mass scales self-consistently.   All CDM simulations have a finite mass resolution which is likely very far from the mass of the smallest clump and exceedingly far from the mass of any hypothetical DM particle.
As one immediately spots  in Eq.~(\ref{eq:trelax}), $t_{\rm relax}$ drops with decreasing number of particles $N$, therefore, it is conceivable that a few massive unresolved DM clumps can thermalize a self-gravitating system even if the underlying DM particles are tiny.   The question arises as to whether two-body collisions between unresolved clumps are efficient enough. 

Here we work out the effective cross section resulting from the unresolved clump-clump collisions assuming the clumps to be point masses. We follow the argument given by \citet[][Section~1.2.1]{2008gady.book.....B}, expanded to consider a full distribution of masses rather than a single mass.

Each single collision with a mass $\mu$ changes the velocity of the collider $\upsilon$ by
\begin{equation}
  \delta \upsilon \simeq \frac{2G\mu}{l\,\upsilon},
  \label{eq:appa}
\end{equation}
where $l$ stands for the impact parameter, i.e., the minimum distance between target and collider. $G$ is the gravitational constant. In a medium with a number density of particles per unit mass $P(\mu)$, the number of collisions during the time interval $d t$, with targets having impact parameter $l\pm dl/2$ and masses in the interval $d\mu$ is 
\begin{equation}
  P(\mu)d\mu\, \upsilon dt\, 2\pi ldl.
  \label{eq:appb}
\end{equation}
By definition, the relaxation timescale $t_{\rm relax}$ is the time span needed for the collective action of all the individual collisions to modify the velocity in as much as $\upsilon$. Considering the change produced in a single collision (Eq.~[\ref{eq:appa}]), the number of collisions per unit time (Eq.~[\ref{eq:appb}]), and the fact that different collitions are random and independent, $t_{\rm relax}$ is implicitly defined as
\begin{equation}
2\pi \int_0^{t_{\rm relax}}\Big[\int_{\forall \mu}\int_{\forall l}\,\upsilon\,P(\mu)\,(\delta \upsilon)^2 \,l dl\,d\mu\Big]\,dt\simeq \upsilon^2.
  \label{eq:appc}
\end{equation}
Using Eq.~(\ref{eq:appa}), it is clear that the integral on the impact parameter in Eq.~(\ref{eq:appc}) is a natural logarithm, i.e.,
\begin{equation}
 \int_{\forall l}\,(\delta \upsilon)^2 \,l dl\propto  \int_{\forall l}\,l^{-1}\, dl=\ln(l_{\rm max}/l_{\rm min})=\ln[M/(2\mu)].
  \label{eq:appd}
  \end{equation}
  The symbols $l_{\rm max}$ and $l_{\rm min}$ stand for the largest and smallest impact parameter, respectively. 
  In the last identity of Eq.~(\ref{eq:appd}), $l_{\rm max}$ is replaced by the radius of the self-gravitating system $R$ and $l_{\rm min}$ is set by the impact parameter corresponding to a strong collision producing $(\delta \upsilon)^2 \simeq \upsilon^2$, i.e., when Eq.~(\ref{eq:appa}) starts to be invalid and the relaxation is reached in a single collision. We have also assumed that the typical velocity of the collider is set by the gravitational well, thus,
  \begin{equation}
    \upsilon^2\simeq \frac{G\,M}{R}, 
  \label{eq:appe}
    \end{equation}
with $M$ the total mass of the system. 
The  cross-section is defined in terms of the relaxation timescale in Eq.~(\ref{eq:sidm1}),
\begin{equation}
  \sigma_{\rm relax}\equiv \frac{1}{P_0\, \upsilon\,t_{\rm relax}},
\label{eq:appsigma}
\end{equation}
with $P_0$ the number density of particles to collide with,
\begin{equation}
 P_0 = \int_{\forall \mu}\,P(\mu)\,d\mu.
\end{equation}
Equations~(\ref{eq:appc}), (\ref{eq:appd}), (\ref{eq:appe}), and (\ref{eq:appsigma}) lead to,
  \begin{equation}
    \sigma_{\rm relax} =\frac{8\pi R^2}{M^2}\,\Big[I_1\,\ln\frac{M}{2} -I_2\Big],
    \label{app:main1}
  \end{equation}
  \begin{equation}
I_1 =\int_{m_p}^{m_{\rm resol}}\mu^2\,f(\mu)\,d\mu,
    \end{equation}
  \begin{equation}
I_2 =\int_{m_p}^{m_{\rm resol}}\mu^2\,\ln(\mu)\,f(\mu)\,d\mu,
\end{equation}
so that $m_p\le \mu\le m_{\rm resol}$ with $m_p$ the true mass of the DM particle and $m_{\rm resol}$ the mass resolution of the simulation.  The new symbol $f(\mu)$ represents the normalized $P(\mu)$,
\begin{equation}
  f(\mu)=P(\mu)/P_0,
\end{equation}
and it stands for the mass function of clumps, i.e., the probability density function of having an unresolved clump of mass $\mu$. Thus, the mean mass of the unresolved clump $\langle\mu \rangle$ is just
\begin{equation}
  \langle \mu \rangle=\int_{m_p}^{m_{\rm resol}}\,\mu\,f(\mu)\,d\mu.
  \label{app:main2}
\end{equation}

With quite some generality, one can assume $f(\mu)$ to follow a power law for the masses in the range of interest, i.e.,
\begin{equation}
      f(\mu) \propto 
    \begin{cases}
    \mu^{-\beta} & m_{p}\le \mu \le m_{\rm resol},\\
    0 & {\rm elsewhere}.
   \end{cases}
\label{eq:powerlaw}
\end{equation}
This is the low-mass end dependence in the Press-Schechter formalism  \citep[][]{1974ApJ...187..425P}, with the exponent $\beta=1.5$ in the  case of a scale-free power spectrum of primordial fluctuations, which is the spectrum within the standard cosmological model. The same kind of law is found in numerical simulations of rather massive DM haloes \citep[$>10^{11} {\rm M}_\odot$; ][]{2008ApJ...688..709T}, and also in DM halos down to Earth-mass haloes
\citep[e.g.,][]{2008Natur.454..735D,2012PDU.....1...50K}. For the sake of reference, the observed luminosity function of galaxies has an exponent $\beta\lesssim 0.5$ at the low-luminosity end \citep[][]{2005ApJ...631..208B,2009ApJ...695..900Y,2017ApJ...835..159S}.

\begin{figure}
  \begin{center}
  \includegraphics[width=.9\columnwidth]{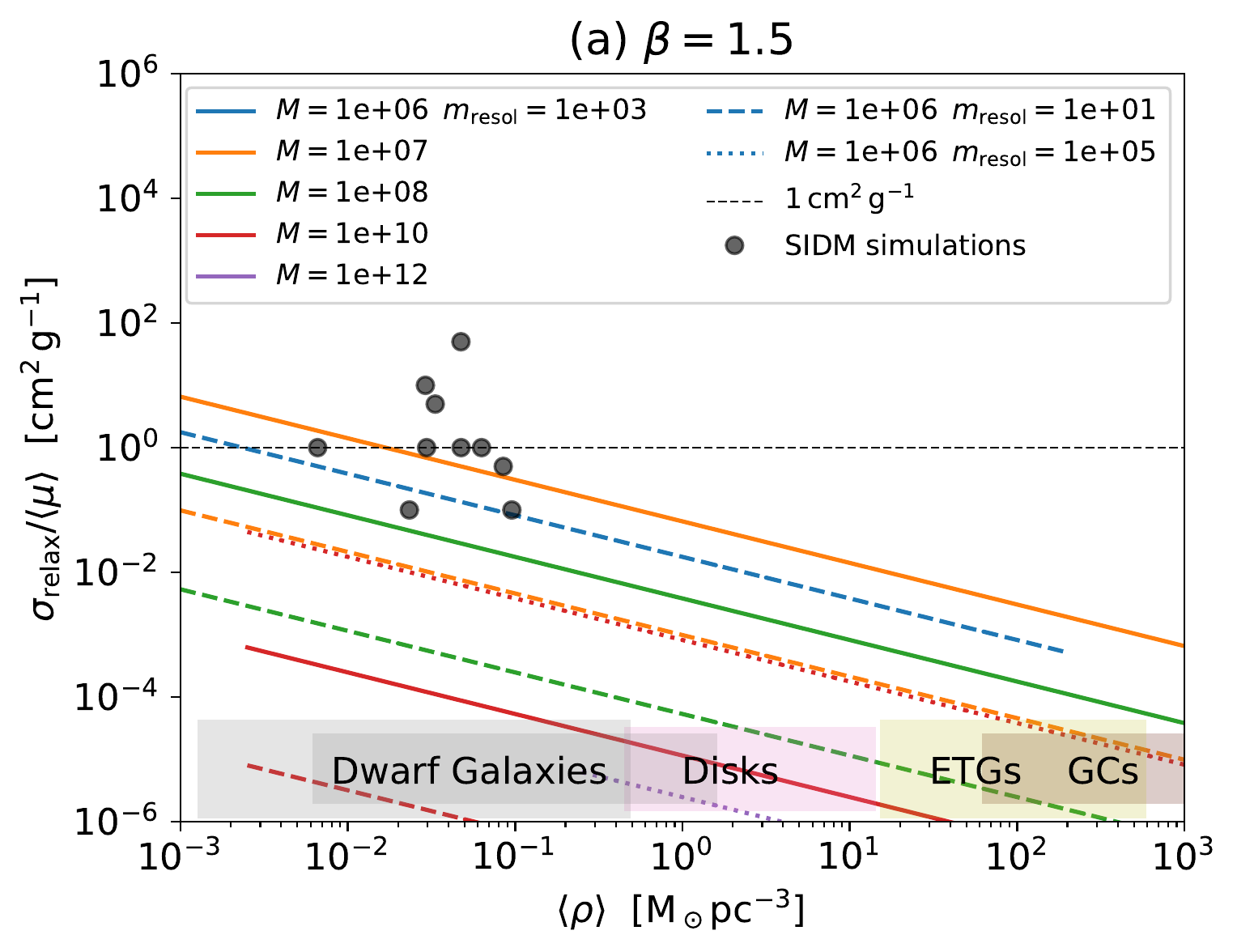}
  \includegraphics[width=.9\columnwidth]{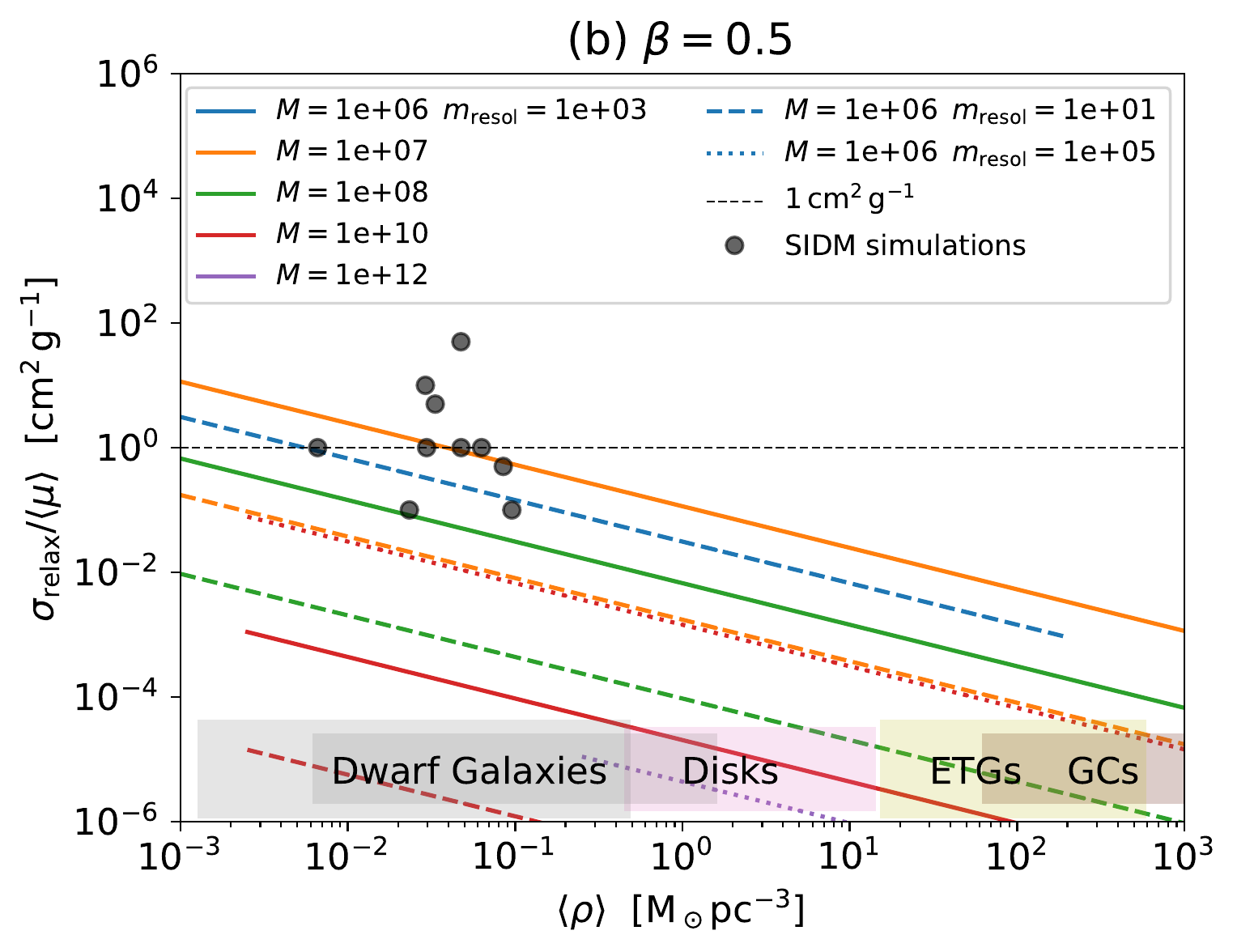}
  \includegraphics[width=0.9\columnwidth]{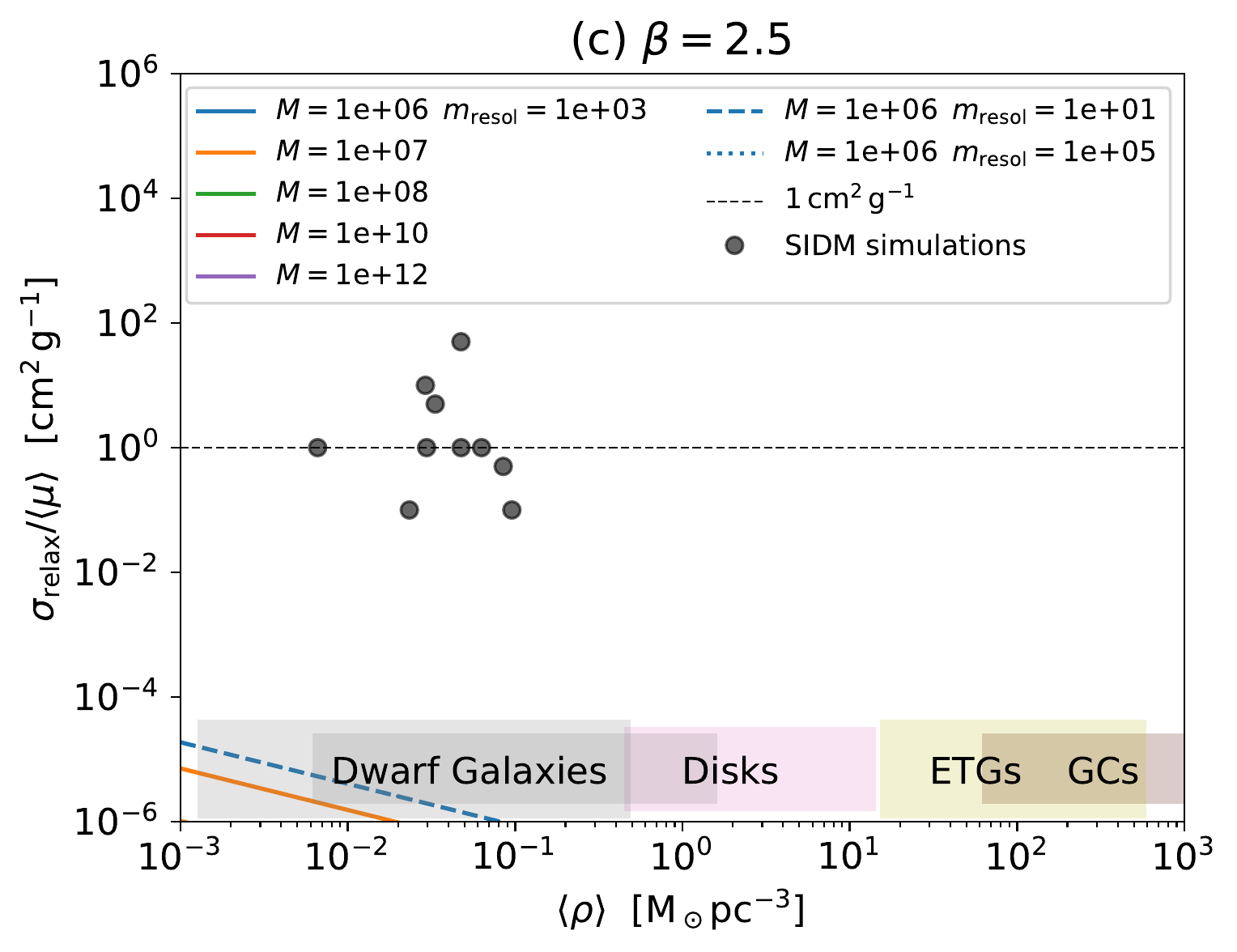}
  \end{center}
  \caption{Dependence of $\sigma_{\rm relax}/\langle\mu \rangle$ on the mean density of the self-gravitating structure modeled with a finite mass resolution $m_{\rm resol}$.  Each line corresponds to a fixed total mass $M$, with each color denoting structures with the same $M$. Different types of lines indicate different mass resolutions in the simulation as labeled in the inset. (Only blue lines are included in the inset but the code is the same for the other colors.) The figures scan a reasonable range of free parameters, namely, $6\le \log(M/{\rm M}_\odot)\le 12$, 10\,pc\,$\le R \le$\,10\,kpc,  $1\le \log(m_{\rm resol}/{\rm M}_\odot)\le 5$, and $0.5\le \beta \le 2.5$. Within this range of $\beta$, $m_p$ plays no significant role and was set to $10^{-10}\, {\rm M}_\odot$.  Only systems with more than 1000 resolution elements in each structure are included.  Panels  (a), (b), and (c) differ only in $\beta$, as indicated in the overhead labels. Masses are given in ${\rm M}_\odot$. The dashed line represents the divide at which numerical simulations of SIDM form a central core. Most lines lie below this divide implying that the existence of unresolved DM clumps do not provide sufficient cross section to explain the existence of cores. The boxes with typical central densities of various astronomical self-gravitating systems are the same as in Fig.~\ref{fig:trelax}, and we refer to the caption of this other figure for further details.}
  \label{fig:timescale5}
\end{figure}

All in all, the effective cross section per unit mass ($\sigma_{\rm relax}/\langle\mu\rangle$), defined by Eqs.~(\ref{app:main1}) and (\ref{app:main2}), depends on 5 parameters, namely, the total mass and size of the self-gravitating system ($M$ and $R$), the mass resolution of the numerical simulation ($m_{\rm resol}$), the mass of the DM particle ($m_p$), and the exponent of the power law characterizing the mass function of the unresolved clumps ($\beta$). Figure~\ref{fig:timescale5} shows that resulting cross section for a physically reasonable range of parameters, namely, $\,6\le \log(M/{\rm M}_\odot)\le 12$, 10\,pc\,$\le R \le$\,10\,kpc,  $1\le \log(m_{\rm resol}/{\rm M}_\odot)\le 5$, and $0.5\le \beta \le 2.5$. Within this range of $\beta$, $m_p$ plays no significant role and was set to $10^{-10}\, {\rm M}_\odot$. The different lines in the figure present the dependence of the cross section on mean density of the system when $M$ is constant. The number of point masses in the numerical simulation is of the order of $M/m_{\rm resol}$, and only systems with this number larger than 1000 are included.  
The dashed line represents the divide at which SIDM numerical simulations form a central core (Section~\ref{sect:SIDMhalo1}). There is a main consequences to be drawn. Most lines lie below this divide implying that the existence of unresolved DM clumps do not provide sufficient cross section to create cores.
The only line the lies above corresponds to DM haloes having $N\simeq 10^{3-4}$, i.e., at the limit of being reliable in numerical simulations.

The decrease of $\sigma_{\rm relax}$ with increasing density is counterintuitive, but it has a simple explanation. The effect of each two-body collision scales with the inverse velocity ($|\delta \upsilon/\upsilon|\propto \upsilon^{-2}$; Eq.~[\ref{eq:appa}]), i.e., slow collisions are far more effective than fast collisions. The relative velocity between particles increases with increasing density, and this effect overcomes the increase of collision frequency with increasing density. Given $M$, $m_{\rm resol}$, $m_p$, and $\beta$, Eq.~(\ref{app:main1}) predicts a power law drop of cross section with average density, explicitly, 
\begin{equation}
  \sigma_{\rm relax}\propto \langle\rho\rangle^{-2/3}.
  \end{equation}


\bsp	
\label{lastpage}
\end{document}